\def\be{\begin{equation}}
\def\ee{\end{equation}}

\documentstyle[twocolumn,prd,aps]{revtex}

\input epsf

\begin{document}
\draft

\title{Instability of an ``Approximate Black Hole"}   

\author{Matthew W. Choptuik, Eric W. Hirschmann, and Steven L. Liebling}

\address{Center for Relativity, The University of Texas at Austin, 
Austin, TX 78712-1081}

\maketitle

\begin{abstract}
We investigate the stability of a family of spherically symmetric static solutions in vacuum 
Brans-Dicke theory (with $\omega=0$) recently described by van Putten.
Using linear perturbation theory,
we find one exponentially growing mode for every member
of the family of solutions, and thus conclude that the solutions are not
stable. 
Using a previously
constructed code for spherically symmetric Brans-Dicke,
additional evidence for instability is provided by directly evolving
the static solutions with perturbations.
The full non-linear evolutions also suggest that the solutions 
are black-hole-threshold critical solutions.
\end{abstract}

   
Recently van~Putten has proposed a one parameter family (parameter $\epsilon$) of 
 solutions to spherically symmetric Brans-Dicke theory for
use in numerical relativity as an approximate black hole \cite{putten}.
These solutions have the attractive property that for small values of this   
parameter, 
the ``exterior" solution approaches that of Schwarzschild. 
However, the event horizon of Schwarzschild is replaced with a high red-shift
horizon and all metric components remain finite at this horizon.
In addition to regularizing the horizon, these solutions have a global timelike
coordinate.  

As van~Putten stresses, these solutions could have 
promise for numerical relativity
because of the difficulties that arise when dealing numerically with
boundary conditions at the horizon of a black hole.  
To be useful as approximate black holes, however, the solutions, like Schwarzschild, must be stable.

In this Communication, we recompute the solutions considered by van~Putten
and carry out a  linear
perturbation analysis about them.  In so doing, we find, for generic
$\epsilon$, modes which grow exponentially in time.  We also directly evolve 
perturbations on the background of these solutions 
and confirm the instability predicted by linear theory.
Thus although of some theoretical interest, these solutions are unlikely 
to be of direct use in the context of mocking-up a black hole in general relativity.

To begin, let us review the static 
family of solutions considered by van~Putten.  
We note that they were 
first 
written down by Brans and Dicke\cite{bd} but in an isotropic 
coordinate system as opposed to the Schwarzschild-like coordinates that
van~Putten uses.  

We work in Brans-Dicke theory and assume spherical 
symmetry.  We choose a coordinate system such that 
the metric has the form
\be
ds^2 = 
         -e^{\nu(r,t)} dt^2 + e^{\lambda(r,t)} dr^2 
        + r^2 d\Omega^2.
\ee
Using $\phi(r,t)$ for the Brans-Dicke field, the field equations are \cite{bd}
\be
G_{\mu \nu} = \frac{8 \pi}{\phi}T_{\mu \nu},
\ee
where, in vacuum, the Brans-Dicke stress tensor is given by  
\begin{eqnarray}
T_{\mu \nu} & = & \frac{\omega}{8 \pi \phi} \left(
        \phi_{,\mu} \phi_{,\nu} -\frac{1}{2} g_{\mu \nu}
        \phi_{,\rho} \phi^{,\rho} \right) 
\nonumber \\
  & & + \frac{1}{8\pi} \left(\phi_{;\mu \nu} -
        g_{\mu \nu} \Box \phi \right),
\label{eq:stress}
\end{eqnarray}
and $\omega$ is the Brans-Dicke coupling constant \cite{weinberg}.
The field $\phi$ satisfies the covariant
wave equation \cite{weinberg}
\be
\Box \phi = \frac{8\pi }{2\omega + 3} T^{{\rm matter}} = 0.
\ee
This equation, along with van~Putten's restriction to $\omega=0$,
simplifies Eq.~(\ref{eq:stress}) to $T_{\mu \nu} = \phi_{;\mu \nu}/8 \pi$.
As the stress tensor is traceless, the field equations
may then be written in the simple form
\be
R_{\mu \nu} = \frac{\phi_{;\mu\nu}}{\phi},
\label{eq:rmunu}
\ee
where $R_{\mu \nu}$ is the usual Ricci curvature tensor.  

We introduce a new field $\psi(r,t)$ 
such that
\be
 e^{\psi(r,t)} = \frac{A}{r \phi(r,t) },
\ee
where $A$ is a constant \cite{eric}.
The field equations are
\begin{eqnarray}
{\dot{\lambda} \over r} + \dot{\psi}' & = & (\dot{\psi}+{\dot{\lambda}\over 2})
    ({1\over r} + \psi')
     +  {1\over 2} \nu' \dot{\psi} \nonumber \\
 -{e^{\lambda} \over r} & = & \psi' + {1\over2}(\lambda' - \nu') 
\label{eq:total} \\
\left[ e^{{1\over2}(\lambda-\nu) -\psi} r \dot{\psi} \right]^{\dot{}}
  & = &   \left[ e^{{1\over2}(\nu-\lambda) -\psi}(1 + r\psi') \right]' \nonumber\\ 
      e^{\lambda-\nu}(2 \ddot{\psi} 
    - \dot\psi(2\dot{\psi} + \dot{\nu}) )
&  = & 
({1\over r} +\psi')(\nu' + {2\over r})
-{1\over r}(\lambda'+\nu'), 
	 \nonumber  
\end{eqnarray}
where overdots and primes
denote derivatives with respect to $t$ and $r$, respectively.  
The first three equations above correspond to the $tr$ and
$\theta\theta$ components of Eq. (\ref{eq:rmunu}),
and the wave equation, respectively.  The final
equation is a convenient linear 
combination of the $tt$ and  $rr$ components of Eq. (\ref{eq:rmunu}),
and the wave equation.

To find the time-independent solutions as van~Putten does, all
time derivatives appearing in Eqs.~(\ref{eq:total}) are set to zero, yielding
$$
   \psi'+ {1\over2}(\lambda' - \nu' )  =  {-e^{\lambda}\over r}  
                {\rm ~~~~~~~~} 
   \psi'  =  {1\over r} (2 Z - 1) 
$$
\be
\lambda' + \nu'  =  2 Z ({2\over r} + \nu'),   \label{eq:unp}
\ee
where $Z \equiv \frac{1}{2}e^{\psi+\frac{1}{2}\left( \lambda - \nu \right)}$
and
is identical  to that defined by van~Putten \cite{putten}.
Details regarding the solution of this system of equations can be found in \cite{putten};
here we will only quote the results.
The metric components are 
\begin{eqnarray} 
e^{\lambda} & = & 1 - 4Z + \frac{1}{\epsilon}Z\left(1 - Z\right)
   \equiv  {1\over\epsilon} (Z_{1}+Z) (Z_{2}-Z) \nonumber\\  
e^{\nu} & = & \left[ {1 - {Z\over Z_{2}} \over 1 + {Z\over Z_{1}}}\right]
     ^{1\over Z_{1} + Z_{2}},  \nonumber  
\end{eqnarray}
where $\epsilon$ is an integration constant, and the constants $Z_1$ and $Z_2$
are given by  
\begin{eqnarray}
Z_1 &= & 2\epsilon-{1\over2} + {1\over 2} \sqrt{1-4\epsilon+16\epsilon^2}
\nonumber \\
Z_2 &= & -2\epsilon+{1\over2} + {1\over 2} \sqrt{1-4\epsilon+16\epsilon^2}.
\end{eqnarray}
The field $Z$ is found  from the transcendental equation
\be
{|Z|^{Z_{1}+Z_{2}} \over |Z_{2}-Z|^{Z_{1}} |Z+Z_{1}|^{Z_{2}}} = 
     r^{-(Z_1 + Z_2)}. 
\label{eq:trans}
\ee
Note that the field $\psi$ can be recovered once $Z$, $\lambda$, and $\nu$ are known.  
 
As van Putten points out, Eq.~(\ref{eq:trans}) has four solutions,
only one of which is Schwarzschild-like in its exterior (van Putten's
type Ia~\cite{putten}). For this solution we have  
$\epsilon>0$ and $Z\rightarrow Z_{2}$ as $r\rightarrow0$ while 
$Z\rightarrow0$ as $r\rightarrow\infty$.

It is worthwhile to consider the small $r$ behavior of these fields.  
In terms of the  
integration constant $\epsilon$, this behavior is found to be
\begin{eqnarray}
e^\lambda & \approx & {Z_2^{2} \over \epsilon} \left( { Z_2 \over Z_1 + Z_2}
    \right)^{{Z_2 \over Z_1} - 1} r^{{Z_2 \over \epsilon} (Z_1 + Z_2)} 
\nonumber \\
e^\nu & \approx & \left( { Z_2 \over Z_1 + Z_2}
                   \right)^{{Z_2 \over Z_1} {1 \over Z_1 + Z_2}} 
    \left( \frac{Z_1}{Z_1 + Z_2} \right)^{\frac{1}{Z_1 + Z_2}} 
    r^{{Z_2\over\epsilon}}.
\label{eq:init} 
\end{eqnarray}
If in addition to small $r$, we consider the limit of small $\epsilon$,
these expressions reduce to 
\begin{equation}
e^{\lambda}  \approx  {1 \over e\epsilon} r^{1/\epsilon - 5} 
     {\rm ~~~~~~~}
e^{\nu}  \approx  {\epsilon \over e} r^{1/\epsilon - 3}. 
\end{equation}

\begin{figure}
\epsfxsize=7cm
\centerline{\epsffile{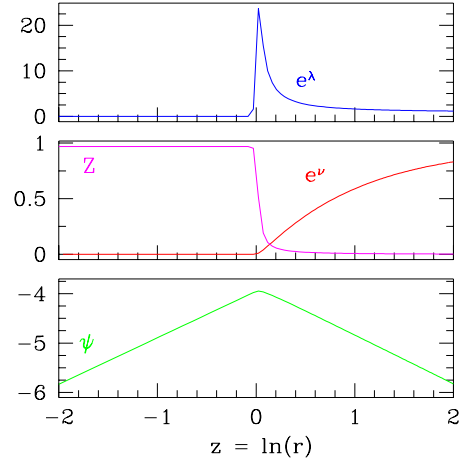}}
\caption{Unperturbed fields for $\epsilon = 0.01$. 
As $\epsilon\rightarrow 0$,
the field $Z$ approaches a step-function (see Fig.~\ref{fig:zos}) and the
field $e^\lambda$ becomes more sharply peaked.
This value of $\epsilon$ is chosen to 
correspond with Fig. 2 of [1].}
\label{fig:eps0.01}
\end{figure}

Fig.~\ref{fig:eps0.01} displays the solution to Eqs.~(\ref{eq:unp})
subject to the initial conditions derived from (\ref{eq:init}) 
for $\epsilon = 1/100$ (the
same value shown in \cite{putten}).

Having constructed these static solutions, we can now address 
the question of their stability.   
For a given $\epsilon$, such a time-independent solution used in
numerical relativity as an approximate black hole, were it not stable,
would either collapse to a black hole or possibly disperse leaving flat space.

To investigate stability, we perform a standard linear perturbation analysis.
As such, we consider the case that the fields {\em do} possess
a small time dependent part  and make the following expansion
for small $\delta$
\begin{eqnarray}
\psi    & \rightarrow & \psi_0(r)    + \delta \hspace{0.02in}\psi_1(r,t)    \nonumber \\
\nu     & \rightarrow & \nu_0(r)     + \delta \hspace{0.02in}\nu_1(r,t) \label{eq:tran}    \\
\lambda & \rightarrow & \lambda_0(r) + \delta \hspace{0.02in}\lambda_1(r,t), \nonumber
\end{eqnarray}
where subscripts $0$ and $1$ denote the unperturbed and perturbed fields,
respectively.

We substitute
the expansion (\ref{eq:tran}) into the full set of Eqs.~(\ref{eq:total}), 
keeping only terms to linear order.
Because the unperturbed fields
satisfy (\ref{eq:total}) by construction, we are left
with the following linear equations for ($\psi_1$, $\nu_1$, $\lambda_1$)
\begin{eqnarray}
{\dot{\lambda_1} \over r} & = & (\dot{\psi_1}+{\dot{\lambda_1}\over 2})
    ({1\over r} + \psi_0 ') - \dot{\psi_1}'
     +  {1\over 2} \nu_0 ' \dot{\psi_1}  \nonumber \\
-{e^{\lambda_0} \over r} \lambda_1 & = & \psi_1 ' + {1\over2}(\lambda_1 ' - \nu_1 ') 
      \label{eq:perturbed}\\
0 & = & \psi_1 '' + ({1\over r} + \psi_0') \left[{1\over2}(\nu_1 ' -
    \lambda_1 ') - \psi_1 ' \right] 
\nonumber \\
  & &  + \psi_1 ' \left[ {1\over2}(\nu_0 ' - \lambda_0 ') - \psi_0 ' +
         {1\over r} \right]
    - e^{\lambda_0 - \nu_0} \ddot{\psi_1}  \nonumber\\  
{1\over r}(\lambda_1 '+\nu_1 ') & = & ({1\over r} +\psi_0 ')\nu_1 '
     + \psi_1 ' (\nu_0 ' + {2\over r} )
     -  2 e^{\lambda_0 - \nu_0}\ddot{\psi_1}. \nonumber  
\end{eqnarray}

We perform the standard Fourier decomposition 
of $\psi_{1}(t,r)$  
\be
\hat{\psi}_1(r,\sigma) =  \int e^{i\sigma t} \psi_1(r,t) dt
\label{eq:four}
\ee
and the other perturbed fields.
On substitution into (\ref{eq:perturbed}), 
the defining relation for $\hat{\psi}_1$ then decouples from $\hat{\lambda}_1$
and $\hat{\nu}_1$, yielding a single equation 
\begin{eqnarray}
0 & = & \hat \psi_1 '' + \hat \psi_1 ' {1\over r} \left[ 1 + e^{\lambda_0} {3Z_0 - 1 \over
           Z_0 - 1} \right]  \nonumber \\
  &   & + \hat \psi_1 \left[ \sigma^2 e^{\lambda_0 - \nu_0} -
    {Z_0 \over r (Z_0 - 1)}
    (\nu_0 ' + {4 Z_0 \over r}) e^{\lambda_0}\right]  
\label{eq:pert}
\end{eqnarray}
which can be solved for the mode $\hat \psi_1(r)$.
We have thus reduced the perturbation problem to a single  
one-dimensional ODE with the undetermined 
eigenvalue $\sigma^2$.

\begin{figure}
\epsfxsize=7cm
\centerline{\epsffile{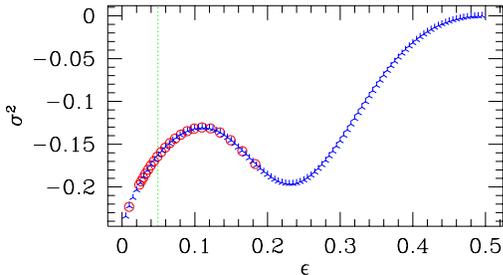}}
\vspace{-1.2in}
\caption{Plot of the unstable eigenvalues $\sigma^2$ versus parameter $\epsilon$.
The cross-marks
and open circles represent data from two independent evaluations of the modes.
Their correspondence indicates a quite small uncertainty. The vertical dotted
line denotes the value of $\epsilon$ for which 
Figs.~\ref{fig:evolve},~\ref{fig:modes},~and~\ref{fig:critical} are computed.
}
\label{fig:sigmas}
\end{figure}

Solution of Eq.~(\ref{eq:pert}) requires appropriate boundary conditions on
$\hat \psi_1(r)$.  
It is common to enforce regularity at the origin of a spherically symmetric spacetime,
but in the current case
the unperturbed solution is
itself irregular.  
Thus, assuming regularity of the perturbation
might seem improper.  However, although $\psi_0$ is logarithmically
divergent 
at $r=0$,  
$\exp(\psi_0)$ is regular at the origin, going to zero as a positive power
of $r$.  
Hence,  it is not unreasonable
to impose regularity on the field $e^{\psi}$. 
Further, because $\exp(\psi) = \exp(\psi_0+\delta\psi_1) = \exp(\psi_0)(1+\delta\psi_1)$,
it is reasonable to assume the regularity of $\hat \psi_1$ at the origin. At most,
the mode $\hat \psi_1$ could have a logarithmic divergence, however, if we can
find an unstable mode with the stricter criterion of regularity, then the solutions
are still, in general, unstable.

Enforcing the assumption of regularity of $\hat{\psi}_1$, allows
us to find a series expansion for $\hat{\psi}_1$ near the origin.  
For very small $r$, Eq.~(\ref{eq:pert}) becomes 
\be
0 = \hat{\psi}_1'' + \hat{\psi}_1' {1\over r} + \sigma^2 C \hat{\psi}_1 r^{p}.
\ee
The positive coefficient $C$ is determined from Eq.~(\ref{eq:init}) 
and depends on
$\epsilon$.  The exponent $p = {Z_2 \over \epsilon}(Z_1 + Z_2 - 1)$ 
likewise depends only on $\epsilon$ and in such a way that 
$p>-2$ for $\epsilon>0$.  We can now find an expansion for $r\ll 1$ 
$$
\hat{\psi}_1 (r) = \hat{\psi}_1(0) \left[1 - {\sigma^2 C \over 2+p} 
      r^{2+p} + \left({\sigma^2 C \over 2+p}\right)^2 r^{4+2p} 
       + \ldots\right] .
$$
Because of the linearity of the problem, $\hat{\psi}_1(0)$ can be
arbitrarily chosen as it reflects the scaling in the problem.
It turns out that we need to use only the first couple of terms
in the expansion to get accurate results.

\begin{figure}
\epsfxsize=7cm
\centerline{\epsffile{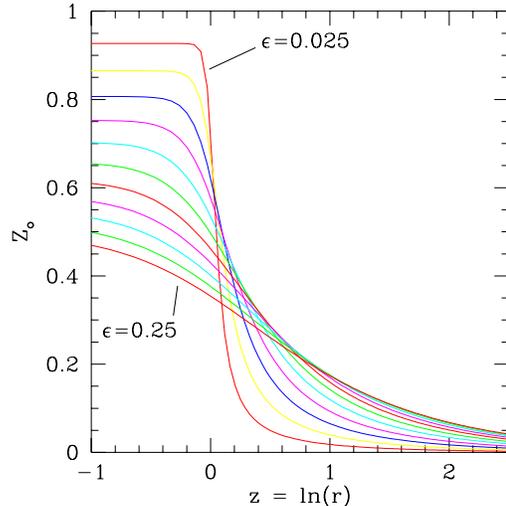}}
\caption{The unperturbed field $Z_0(r)$ 
for $0.025 \le \epsilon \le 0.24$ (uniform in $\epsilon$).
The field $Z_0(r)$ approaches a step function as $\epsilon$ is decreased.
The approach of the unperturbed $Z_0$ to zero at large $z$ signifies
the approach to Schwarzschild in that region. For $z<0$, however,
the solution clearly is not Schwarzschild.
}
\label{fig:zos}
\end{figure}

Given a background solution to Eqs.~(\ref{eq:unp})
for a particular $\epsilon$, we can now 
solve the eigenvalue problem (\ref{eq:pert}) for the modes
$\hat \psi_1(r)$ and corresponding characteristic frequencies $\sigma^2$.
In our  
particular case, the instability of the original soliton solutions 
is indicated by the existence of one or 
more 
exponentially growing modes.  These are solutions to the perturbation 
equations with negative eigenvalues:  
$\sigma^2<0$.  

In practice, we integrate the unperturbed equations and the 
perturbation equation simultaneously from $r\approx 0$ to large $r$,
looking for a solution which has a negative eigenvalue and obeys
the boundary conditions. Specifically, we demand that the mode
be finite at the origin and vanish asymptotically. 
We use a
standard ODE integrator 
and standard shooting techniques in our search.

Although our search has not been exhaustive, 
we generically find precisely one growing mode for each value of $\epsilon$.
This is sufficient for us to conclude that
the static  solutions  are unstable.
The eigenvalues found for these solutions
are shown in Fig.~\ref{fig:sigmas}.

Having found the perturbation modes, looking at the limit $\epsilon\rightarrow0$
is instructive. As this limit is approached, the unperturbed solution
becomes more and more like Schwarzschild in the exterior, 
and this resemblance is
precisely the reason why  the family  has been proposed as a good model of a black hole. 
With this in mind, one may wonder why Fig.~\ref{fig:sigmas} shows
that, as $\epsilon \rightarrow 0$, there is still a growing mode. Certainly
these results do not show Schwarzschild to be unstable; rather
we point out that within the ``effective horizon'' of this
approximate black hole, the solution is very different from the interior 
Schwarzschild solution for all 
$\epsilon$ (see Fig.~\ref{fig:zos}).  
Hence, it is reasonable to assert that the solution is
unstable for any $\epsilon$, including the solution $\epsilon=0$.

\begin{figure}
\epsfxsize=7cm
\centerline{\epsffile{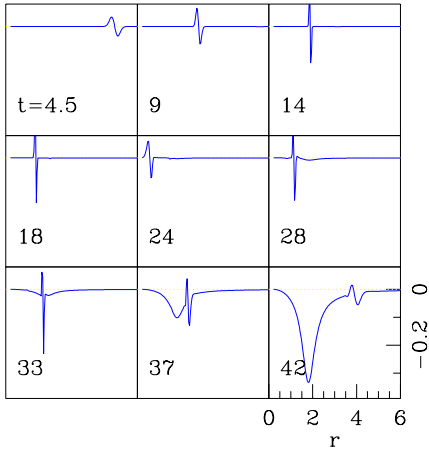}}
\caption{Series of snapshots of
the product $r\Pi_\xi$ (related to the time derivative of $\psi$) 
in the Einstein frame.  A
small perturbation ($\Delta M/M =0.018\%$) at large radius is introduced to the
initially static solution ($\epsilon = 0.127$).
The perturbation passes through the singularity at $r=0$ (between
$t=24$ and $28$) and escapes to $r=\infty$.
As the perturbation passes the red-shift horizon (as it propagates inwards, the
perturbation is seen to experience a blue-shift), the
excitation of a growing mode is clearly seen.
}
\label{fig:evolve}
\end{figure}

Further, since it is the case that to an outside observer, 
the $\epsilon=0$ solution
is indistinguishable from that of Schwarzschild, it is logical to 
assume that any perturbation of the solution will not change the
view of this observer. In other words, as $\epsilon\rightarrow 0$, 
any perturbation
should have decreasing support outside the ``effective horizon" 
of the approximate black hole.
Indeed, we observe this kind of behavior. As one decreases
$\epsilon$, the profile  
of the mode $\hat \psi_1$ is seen to approach that of a delta function  
at the position
where the apparent horizon would asymptotically form.

We find additional evidence for instability by evolving 
the static solutions (with small perturbations added) 
using the full time-dependent equations of motion \cite{liebling}.
This has the added benefit of providing information concerning 
the states to which the static solutions evolve when perturbed.
To this end, we have adapted a previously developed  spherically symmetric
code for Brans-Dicke theory~\cite{liebling} which allows
us to evolve these solutions. Because the evolutions 
in \cite{liebling} are performed 
in the Einstein conformal frame (as compared to the Brans-Dicke frame in 
which van Putten works and in which we have worked thus far), 
we transform the fields used above to the Einstein
frame.   In this frame, the field equations are equivalent to those 
for a massless scalar field minimally coupled to gravity.

\begin{figure}
\epsfysize=7cm
\centerline{\epsffile{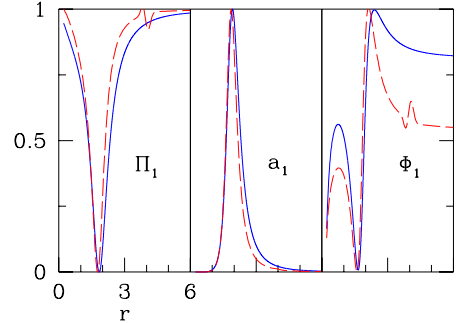}}
\vspace{-0.65in}
\caption{
Comparison of modes computed in perturbation theory (solid lines)
with modes computed from the full non-linear evolution (dashed lines).
The non-linear modes were computed by taking a late-time profile from the evolution and subtracting
the initial data for that field, as defined in Eq.~(\ref{eq:modes}).
Perturbative modes have been numerically transformed to the Einstein frame.
All fields are rescaled
to the interval $[0..1]$ and are plotted with respect to $r$. 
}
\label{fig:modes}
\end{figure}

After recovering the static solutions in the Einstein frame and
inputing a transformed solution into the code, 
we introduce a small ingoing perturbation to the fields
at large radius.          
For generic values of $\epsilon$, we find that
van Putten's solution either collapses or disperses after the perturbation
reaches the high-red shift horizon. Fig.~\ref{fig:evolve} 
clearly demonstrates the instability for  a specific  $\epsilon$. 
In this case, the perturbation induces
collapse to a black hole.

In order to facilitate comparison of the perturbation results with
those of the fully nonlinear evolution, we first define the quantities
below in terms of the fields found~in~\cite{liebling}
\begin{eqnarray}
\bar{a}_1       & = & a(r,t) - a(r,0) \nonumber \\
 \bar{\Phi}_1   & = & \Phi_\xi(r,t) - \Phi_\xi(r,0)  \label{eq:modes}\\        
 \bar{\Pi}_1    & = & \Pi_\xi(r,t) - \Pi_\xi(r,0)  \nonumber
\end{eqnarray}
such that fields with a bar and a subscripted $1$ denote nonlinear
deviations from the unperturbed solution ({\em nonlinear modes}).
By transforming the modes found from linear perturbation theory to the
Einstein frame, we may now compare directly the linear modes with
the nonlinear modes ($\bar{\Pi}_1$, $\bar{a}_1$, $\bar{\Phi}_1$).
In Fig.~\ref{fig:modes} we show  all the modes rescaled to the unit interval.
From the near congruence, we conclude that we are observing
the actual evolution of the growth of these perturbation modes.

\begin{figure}
\epsfxsize=7cm
\centerline{\epsffile{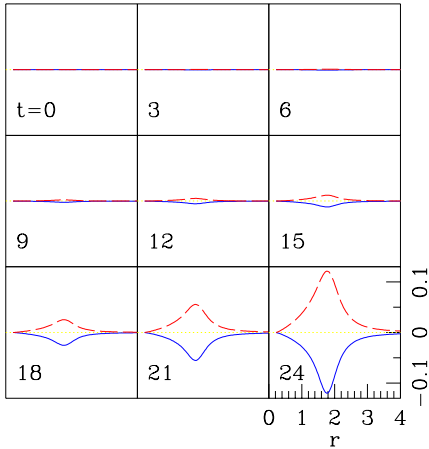}}
\caption{Uniform-in-time series of solution when initially perturbed
with the predicted mode ($\Delta M/M = 0.004\%$, $\epsilon=0.127$).  
Shown here is the field $\Pi_\xi$ (related to the time derivative of $\psi$).
The evolution shown in solid results in collapse to a black hole. The dashed
line shows the evolution resulting from switching the sign of the
perturbation at the initial time. That the initial sign of the introduced
perturbation separates eventual collapse from dispersal indicates
that the unperturbed solution is critical. 
 }
\label{fig:critical}
\end{figure}

In addition to confirming our perturbative results, the full evolution
provides evidence that these static solutions represent critical solutions
to black hole formation. By this we mean that 
these solutions represent a boundary
in the space of solutions between those that form black holes and those
that do not. To demonstrate this criticality 
we begin with an unperturbed solution in the Einstein frame
and add a perturbation to the fields.
In this case instead of the arbitrary perturbation at large $r$ shown 
in Fig. \ref{fig:evolve}, 
we add the predicted 
mode found in the perturbation analysis (solid lines of Fig. \ref{fig:modes}). 
When this mode is added with some small positive amplitude,
we invariably see collapse of the van Putten approximate black hole 
to a genuine black hole
(see the solid lines in Fig. \ref{fig:critical}). In contrast,
when the perturbation is added with a small negative amplitude,
we see dispersal of the solution 
(see the dashed lines in Fig. \ref{fig:critical}). 
Thus it would appear that this solution sits at the threshold  
between solutions that form
black holes and those that disperse \cite{dispersal}.

Having found evidence that these are threshold solutions, we are led to ask
if they are attracting. For them to represent attracting critical solutions 
(intermediate attractors) the unstable (relevant) mode 
which we find must be shown to be 
unique \cite{koike}.
Because both the perturbation analysis and the full evolution appear to indicate
the presence of only a single unstable mode (see Fig. \ref{fig:modes}),
we suspect that these solutions might well represent an intermediate attractor
for black hole formation.
We plan to address this issue in future work.

From the results presented here, these solutions  
would appear to be analagous 
to the
$n=1$ Bartnik-McKinnon (BM) solution in the 
Einstein-Yang-Mills (EYM) system~\cite{bm}. 
This static solution was found 
to be an intermediate attractor in the gravitational collapse of spherically
symmetric $SU(2)$ fields
with one side 
of the threshold being black hole formation and the other dispersion
of the Yang-Mills field \cite{matt}. 

After completion of this work, we became aware of other work which
had considered solutions similar to those examined here. 
Static, spherically symmetric solutions to the minimally
coupled Einstein-Klein-Gordon equations were studied by Buchdahl\cite{buchdahl} and later 
by Wyman\cite{wyman}.  These solutions were written down in the Einstein
frame in contrast to 
van Putten who works in vacuum Brans-Dicke (which is conformally equivalent
to the Einstein-Klein-Gordon system).  
In addition,  
Jetzer and Scialom were able to show that Wyman's
solutions are generically 
unstable by establishing the existence of a negative upper bound 
for the lowest eigenvalue of the perturbation\cite{jetzer}.

This research was supported in part by NSF PHY9310083 and PHY9318152.



\begin{references}
\bibitem{eric}
In general, we have tried to follow
van~Putten's notation fairly closely. However, we found the field $\psi(r,t)$
to be a useful quantity in the perturbation and use it instead of van~Putten's
$\gamma(r,t)$.  Note too that at one point in his derivation van~Putten uses
$\psi$, but for a quantity different from what we have defined above.
\bibitem{dispersal}
What it means for a solution with a naked singularity at the origin to disperse
is perhaps not well-defined at the moment. However, the evolution clearly indicates
that the energy content is moving toward large radius.
\end{references}
\end{document}